\documentclass[prb,twocolumn,showpacs,superscriptaddress,preprintnumbers,amssymb]{revtex4}
\usepackage{graphicx}
\usepackage{latexsym}
\usepackage{amssymb}
\usepackage{amsmath}
\usepackage{amsfonts}
\usepackage{bm}
\usepackage{multirow}
\usepackage{color}
\usepackage{verbatim}
\newcommand{\beq}{\begin{equation}}
\newcommand{\eeq}{\end{equation}}
\newcommand{\beqn}{\begin{eqnarray}}
\newcommand{\eeqn}{\end{eqnarray}}
\newcommand{\al}{\alpha}

\newcommand{\e}{\varepsilon}

\newcommand{\ph}{\varphi}
\newcommand{\da}{\dagger}
\newcommand{\pa}{\partial}

\newcommand{\la}{\mathcal L}

\newcommand{\Ld}{\Lambda}

\newcommand{\ml}{\left(\begin{matrix}}
\newcommand{\mr}{\end{matrix}\right)}

\newcommand{\U}{\mathcal U}

\newcommand{\tr}{\text{tr}}
\newcommand{\op}{\mathcal O}
\newcommand{\del}{\delta}

\newcommand{\half}{\tfrac{1}{2}}

\newcommand{\fourth}{\tfrac{1}{4}}

\newcommand{\eighth}{\tfrac{1}{8}}

\begin{document}

\title{Stable Interacting $(2 + 1)d$ Conformal Field Theories at the Boundary of a class of \\ $(3 + 1)d$
Symmetry Protected Topological Phases}

\author{Zhen Bi}

\author{Alex Rasmussen}

\affiliation{Department of Physics, University of California,
Santa Barbara, CA 93106, USA}

\author{Yoni BenTov}

\affiliation{Institute for Quantum Information and Matter,
California Institute of Technology, Pasadena, CA 91125}

\author{Cenke Xu}

\affiliation{Department of Physics, University of California,
Santa Barbara, CA 93106, USA}

\begin{abstract}

Motivated by recent studies of symmetry protected topological
(SPT) phases, we explore the possible gapless quantum disordered
phases in the $(2+1)d$ nonlinear sigma model defined on the
Grassmannian manifold $\frac{U(N)}{U(n)\times U(N - n)}$ with a
Wess-Zumino-Witten (WZW) term at level $k$, which is the effective
low energy field theory of the boundary of certain $(3+1)d$ SPT
states. With $k = 0$, this model has a well-controlled large-$N$
limit, $i.e.$ its renormalization group equations can be computed
exactly with large-$N$. However, with the WZW term, the large-$N$
and large-$k$ limit alone is not sufficient for a reliable study
of the nature of the quantum disordered phase. We demonstrate that
through a combined large-$N$, large-$k$ and
$\epsilon-$generalization, a {\it stable} fixed point in the
quantum disordered phase can be reliably located in the large$-N$
limit and leading order $\epsilon-$expansion, which corresponds to
a $(2+1)d$ strongly interacting conformal field theory.

\end{abstract}

\date{\today}

\maketitle

\section{Introduction}
%{\it --- Introduction}

A symmetry protected topological (SPT) phase~\cite{wenspt,wenspt2}
must, by definition, have a boundary state with a nontrivial
spectrum when the system including the boundary preserves certain
global symmetries. Many $(2+1)d$ SPT states can be described with
a similar Chern-Simons theory~\cite{luashvin} as the quantum Hall
states, their $(1 + 1)d$ boundary states are therefore relatively
easy to understand. Thus it is more challenging to understand the
$(3+1)d$ SPT states, whose boundary states can have much richer
physics under strong interaction. The following three types of
$(2+1)d$ states may exist at the boundary of a $(3+1)d$ SPT phase:

{\it 1)} An ordered phase that spontaneously breaks the global
symmetry and hence has degenerate ground states;

{\it 2)} A $(2+1)d$ topologically ordered phase with topological
degeneracy;

{\it 3)} A stable gapless phase which is described by a conformal
field theory (CFT).

Possibilities {\it 1} and {\it 2} have both been studied quite
extensively in the last few years, for both fermionic and bosonic
SPT
states~\cite{TI_fidkowski1,TI_fidkowski2,TI_Qi,TI_max,TI_senthil,senthilashvin,xuclass},
but there is little study about the third possibility, except for
the well-known simplest case of noninteracting topological
insulators/superconductors. In this work we explore the third
possibility of SPT phases: a stable $(2+1)d$ {\it interacting}
conformal field theory (CFT) at the boundary of a $(3+1)d$ SPT
state. This CFT should be stable against any symmetry allowed
perturbations, By ``stable" we mean that all perturbations allowed
by symmetry should be irrelevant (in the renormalization group
sense) at this fixed point.
%This
%work will focus on bosonic SPT phases, as unlike the fermionic
%analogues, bosonic SPT phases necessarily require strong
%interactions.

We will take the ``standard" field theory description of $(3+1)d$
bosonic SPT states, which is a nonlinear sigma model (NLSM) with a
$\Theta$-term in the $(3 + 1)d$ bulk spacetime. The value $\Theta
= 2\pi$ corresponds to the stable fixed point of the SPT phase.
This formula was used to describe and classify bosonic SPT states
in Ref.~\onlinecite{senthilashvin,xusenthil,xu3dspt,xuclass}. With
$\Theta = 2\pi$ in the $(3+1)d$ bulk, the $(2+1)d$ boundary is
described by a NLSM with a Wess-Zumino-Witten (WZW) term with
level $k = 1$. In
Ref.~\onlinecite{senthilashvin,xusenthil,xuclass}, the target
space of the NLSM was the four dimensional sphere $S^4$, a WZW
term can be defined based on the fact that the homotopy group
$\pi_4[S^4] = \mathbb{Z}$. Topological phases with the same
anomaly as this field theory under various anisotropies were
discussed thoroughly in Ref.~\onlinecite{senthilashvin,xuclass}.

The presence of a WZW term is known to drastically change the
behavior of the NLSMs in lower dimensions. In particular, in
$(0+1)d$ a WZW term may lead to degenerate ground states; in
$(1+1)d$ a WZW term drives the NLSM towards a conformally
invariant fixed point~\cite{Witten1984,KnizhnikZamolodchikov1984}.
An explicit renormalization group (RG) calculation in $(1+1)d$
demonstrates that this fixed point is stable and occurs at a
finite value of the NLSM coupling constant~\cite{Witten1984}.

However, unlike these $(1+1)d$ analogues, it is difficult to
perform a controlled calculation for NLSMs with a WZW term in
$(2+1)d$. There are two standard controlled RG calculations for
NLSMs in $3d$ Euclidean space-time: (1) Generalizing the
space-time dimensions to $d= 2+\epsilon$, perform an expansion
with ``small" parameter $\epsilon$, and then extrapolate the
result to $\epsilon \rightarrow 1$; (2) Generalizing the target
manifold to $S^N$ with $N \gg 1$, and perform an expansion with
small parameter $1/N$. But both of these standard approaches fail
in present context because of the WZW term. The first method is
questionable in this context because the topological term can only
be defined in an integer number of space-time dimensions. As for
the second method, the fact that $\pi_d[S^N] = 0$ for $d < N$
implies that a naive generalization from $S^4$ to $S^N$ would
completely miss the contribution from the WZW term. An attempt of
calculating the effect of the WZW term in $(2+1)d$ was made in
Ref.~\onlinecite{moonwzw}, but the calculation there was
uncontrolled for precisely the reasons we mentioned above.

However, we suspect that these difficulties may be only technical
in nature. We expect that the WZW term in $(2+1)d$ may still lead
to a stable conformally invariant fixed point at a finite value of
the coupling. This expectation is (indirectly) supported by recent
quantum Monte Carlo simulation on a $2d$ lattice interacting
fermion model, where a continuous quantum phase transition
described by a $(2+1)d$ NLSM with a topological $\Theta-$term was
found, and $\Theta$ was the tuning parameter for this
transition~\cite{kevinQSH,mengQSH2}. The numerical data suggest
that right at $\Theta = \pi$ this theory is a $(2+1)d$ CFT with
gapless bosonic excitations while no gapless fermion excitations.
A field theory with $\Theta = \pi$ can be viewed as another field
theory with a WZW term under symmetry breaking. Thus the results
in Ref.~\onlinecite{kevinQSH,mengQSH2} actually suggest that the
disordered phase of a $(2+1)d$ NLSM with a WZW term can also be a
stable CFT.

Besides these recent progresses, earlier studies of the deconfined
quantum critical point~\cite{deconfine1,deconfine2} also suggested
that a WZW term in a $(2+1)d$ NLSM could lead to a stable CFT. It
was conjectured that the deconfined quantum critical point
corresponds to the quantum disordered phase of the SO(5) NLSM with
a WZW term at level$-1$~\cite{senthilfisher}, and the SO(5)
symmetry could emerge at this CFT.

The goal of this work is to analytically study the effects of the
WZW term on NLSMs in $(2+1)d$ space-time. In section II we first
take a large-$N$ generalization of the boundary field theory of
$(3+1)d$ SPT states which always permits a WZW term in $(2+1)d$
space-time. This theory has a controlled large-$N$ limit without
the WZW term. In section III we first argue that the large-$N$ and
large-$k$ generalization alone is insufficient to provide a
reliable study of the quantum disordered phase, with presence of
the WZW term.
% (although it is
%already sufficient for a controlled calculation for the critical
%exponents at the order-disorder transition, with certain adequate
%scaling between $N$ and $k$).
Then we demonstrate that a combined large-$N$, large-$k$ and
$\epsilon-$generalization enables us to identify a stable fixed
point in the quantum disordered phase, which corresponds to a
$(2+1)d$ interacting CFT. In section IV, we will briefly discuss
the connection of this work to the ``hierarchy problem" in high
energy physics.

\section{Lagrangian and Method}

We would like to find a NLSM with a WZW term that admits a
controlled approximation scheme for evaluating the RG equations
(beta functions). This means that the target space $\mathcal{M}$
should have an acceptable large-$N$ generalization that permits a
WZW term in $(2+1)d$. One example that satisfies these constraints
is the Grassmannian manifold:
\begin{equation}
\mathcal{M}(n,N) = \frac{U(N)}{U(n)\times U(N-n)}\;,
\end{equation}
For any $n \geq 2$, $N \geq n+2$, $\pi_4[\mathcal{M}] =
\mathbb{Z}$ while $\pi_3[\mathcal{M}] = 0$, thus a WZW term can be
defined in $(2+1)d$ for $\mathcal{M}$.
%~\footnote{The Grassmannian
%$\mathcal{M}$ has $\pi_4[\mathcal{M}] = \mathbb{Z}$ and
%$\pi_3[\mathcal{M}] = 0$, thus the WZW term is the only
%topological term.}.
For $n=1$, this manifold is the familiar $\mathrm{CP}^{N-1}$
manifold, and later we will argue that even for $n=1$ a similar
term in the action may also be defined.

The total dimension of $\mathcal{M}$ scales linearly with $N$
instead of $N^2$ with large-$N$ and fixed $n$, thus without the
WZW term, a NLSM defined with target manifold $\mathcal{M}$ does
not have the infinite planar diagram problem that usually occurs
in matrix models. The entire action in (2+1)$d$ Euclidean
space-time that we will study is \beqn && \mathcal{S} = \int
\!d^2x\, d\tau \ \frac{1}{g} \mathrm{tr}(\partial_\mu \mathcal{P}
\partial^\mu \mathcal{P}) \cr\cr &+&\!\! \!\!\int_0^1\! \!\!du\! \int\! d^2x\, d\tau \ \frac{i 2\pi
k}{256\pi^2} \e^{\mu\nu\rho\lambda}\,\mathrm{tr}(\mathcal{\tilde P} \partial_\mu \mathcal{\tilde P}
\partial_\nu \mathcal{\tilde P} \partial_\rho \mathcal{\tilde P}
\partial_\lambda \mathcal{\tilde P}) .\;\;\;\;\;
\label{3dboundary}\eeqn
%%%%
The basic field $\mathcal{P} \in \mathcal{M}(n,N)$ is an $N\times
N$ hermitian matrix and it can be represented in the form \beqn
\mathcal{P} = V \Omega V^\dagger, \;\;\Omega \equiv\! \left(
\begin{array}{cccc}
\mathbf{1}_{n\times n} & \mathbf{0}_{ n \times (N-n)} \\ \\
\mathbf{0}_{(N - n) \times n} & \;\;\mathbf{-1}_{(N-n) \times
(N-n)}
\end{array}
\right) \;\label{parametrization1}\eeqn where $V\in U(N)$. The
matrix $\mathcal P$ satisfies $\mathcal P^\da = \mathcal P$,
$\mathcal P^2 = I$, and $\tr(\mathcal P) = 2n-N$. (When $N = 2n$,
$\tr (\mathcal P) = 0$, and this was the case studied in
Ref.~\onlinecite{xu3dspt}). %Notice that the parametrization in
%Eq.~\ref{parametrization1} is invariant under the local right
%multiplication $V \to Vh$, where $h(x,\tau)$ is any unitary
%$N$-by-$N$ matrix that commutes with $\W$. Also
Note that when $N = 2$ and $n = 1$, $\mathcal{M}(n,N)$ is
$SU(2)/U(1) = S^2$, and $\mathcal{P}$ can always be represented as
$\mathcal{P} = \vec{n}\cdot \vec{\sigma}$, where $\vec{n}$ is a
three component unit vector, and $\vec{\sigma}$ are the Pauli
matrices.

$\mathcal{\tilde P}(\vec{x}, \tau, u)$ is an extension of
$\mathcal{P}(\vec{x}, \tau)$ into the auxiliary fourth dimension
parameterized by $u\in [0,1]$. This extended field satisfies \beqn
\mathcal{\tilde P}(\vec{x}, \tau, 1) = \mathcal{P}(\vec{x}, \tau),
\ \ \ \mathcal{\tilde P}(\vec{x}, \tau, 0) = \Omega. \eeqn For the
$(2+1)d$ boundary physics described by $\mathcal P(x,\tau)$ to be
independent of the chosen extension $\tilde{\mathcal
P}(x,\tau,u)$, the coefficient $k$ must be quantized. This action
Eq.~\ref{3dboundary} obviously has a global $SU(N)$ symmetry:
$\mathcal{P} \rightarrow U^\dagger \mathcal{P} U$, where $U \in
SU(N)$~\footnote{To be more precise, the global symmetry of this
system is PSU($N$)=$SU(N)/Z_N$ = $U(N)/U(1)$. This is because any
configuration of $\mathcal{P}$ does not transform at all under the
U(1) subgroup of U($N$), or the $Z_N$ center of $SU(N)$. For
example, for $N=2$ and $n=1$, the manifold $\mathcal{M}$ is $S^2$,
and a NLSM defined on $S^2$ should have symmetry SO(3) =
SU(2)/$Z_2$.}.

Our general theory Eq.~\ref{3dboundary} has the following
connections with the previously studied theories:

{\it 1)} In order to study $(3+1)d$ bosonic SPT states,
Ref.~\onlinecite{senthilashvin,xuclass} introduced a NLSM with
target space $S^4$. $S^4$ can also be written as a Grassmannian:
$S^4 \sim \frac{Sp(4)}{Sp(2)\times Sp(2)}$. If written in terms of
$\mathcal{P} = \vec{n}\cdot \vec{\Gamma}$ (where $\vec{n}$ is the
five component unit vector introduced in
Ref.~\onlinecite{senthilashvin,xuclass} and $\Gamma^a$ are the
five $4\times 4$ anticommuting Gamma matrices), the topological
term of Eq.~\ref{3dboundary} is precisely the same as the one in
Ref.~\onlinecite{senthilashvin,xuclass}. Thus the field theory of
Ref.~\onlinecite{senthilashvin,xuclass} can be viewed as our model
with $N = 2n = 4$ after breaking the SU(4) down to smaller
symmetries considered therein.

{\it 2)} Ref.~\onlinecite{abanov2} demonstrated that for $n=1$,
the topological term discussed above can be generated by coupling
the CP$^{N-1}$ manifold to $(2+1)d$ Dirac fermions with SU($N$)
symmetry. Ref.~\onlinecite{senthilhe3} used this fact, and derived
the effective field theory for the bosonic sector for $N=2n=2$,
which corresponds to the boundary of the $(3+1)d$ topological
superconductor with symmetry $SU(2) \times \mathcal{T}$
($\mathcal{T}$ being time-reversal). Ref.~\onlinecite{senthilhe3}
also argued that with the full $SU(2) \times \mathcal{T}$
symmetry, this boundary theory cannot be gapped out, which implies
that it could be an interacting CFT. Thus our theory with
large-$N$ and $n=1$ can also be viewed as a formal generalization
of the case studied in Ref.~\onlinecite{senthilhe3}~\footnote{We
do note that for $N=2n=2$, the space-time integral of the
topological term is quantized, $i.e.$ it is the Hopf term, while
for larger $N$ this term is not quantized.}.

Instead of working with Eq.~\ref{3dboundary} directly, we will use
a parametrization that is more easily amenable to a large-$N$
analysis. This parametrization was introduced in
Ref.~\onlinecite{symmetricNLSM,generalizedNLSM}. We define a
collection of $n$ orthonormal complex vectors
%%%%
\begin{equation}\label{eq:constraint}
\{\vec\ph_\al\}_{\al=1,2,...,n}\;,\;\; \vec \ph_\al^{\,\da}
\cdot\vec\ph_\beta = \del_{\al\beta}.
\end{equation}
The order parameter $\mathcal P$ can be written as
\begin{equation}\label{eq:alternate}
\mathcal P_{ij} = 2\sum_{\al\,=\,1}^n \ph^i_{\al}\ph^{j
\dagger}_{\al}-\del^{ij}
\end{equation}
with $i,j = 1,...,N$. This definition is invariant under local
transformations of the form
%%%%
\begin{equation}\label{eq:gauge transformation}
\ph^i_\al \to \ph_\beta^i \,\U_\al^{\;\;\beta}(x)
\end{equation}
with $\U\in U(n)$. Hence the action in terms of the $\ph_\al^i$
will have a $U(n)$ gauge symmetry, under which each $\ph^i$
transforms as a fundamental $n$-dimensional representation (and $i
= 1,...,N$ serves as a flavor label).

Explicitly, we may observe that the quantity \beqn a \equiv -
id\ph^\da\cdot \ph=-i\sum_{i=1}^N d\ph^{i\da}_\al\ph^i_\beta \label{gauge}
\eeqn transforms as a $U(n)$ gauge field. If we then define the
field strength 2-form $f \equiv da-ia\wedge a$, we find
\begin{equation}\label{eq:f wedge f}
\tr\left( \tilde{\mathcal P}\; d\tilde{\mathcal P}\wedge
d\tilde{\mathcal P}\wedge d\tilde{\mathcal P}\wedge
d\tilde{\mathcal P}\right) = -32\;\tr\left( f\wedge f\right)\;.
\end{equation}
The right-hand side of Eq.~\ref{eq:f wedge f} is a total
derivative in (3+1)$d$, and hence its integral can be reduced to
the (2+1)$d$ integral of a local integrand, namely a $U(n)$
Chern-Simons term.

The right hand side of Eq.~\ref{eq:f wedge f} can also be defined
even for $n=1$ (which corresponds to the case with $\mathcal{M} =
\mathrm{CP}^{N-1}$), and the integral of this term on $T^4$ is
quantized, although its integral on $S^4$ is trivial. This is
analogous to the topological response theory $\sim
\vec{E}\cdot\vec{B}$ of $3d$ topological insulator~\cite{qi2008}.

Following Ref.~\onlinecite{symmetricNLSM,generalizedNLSM}, we
block-decompose the $\ph_\al^i$ fields as
\begin{equation}
\ph_\al^{\;\;i} = (\Phi_{\alpha}^{\beta} \ ; \ \ \phi_\al^{I})^t
\end{equation}
where $I = n+1 \cdots N$. Then we can use local $U(n)$
transformations to make the $n$-by-$n$ block Hermitian (fix the
gauge~\cite{generalizedNLSM}): $\Phi = \Phi^\da$, which eliminates
all the continuous gauge degrees of freedom. The constraint
Eq.~\ref{eq:constraint} on $\ph_\al^{\;i}$ now takes the form:
\begin{equation}
\Phi = (I - \phi^\da \cdot \phi)^{1/2} = I - \half \phi^\da \cdot
\phi - \eighth(\phi^\da \cdot \phi)^2+\op(\phi^6)\;.
\end{equation}
Then we find $\tr[\tilde{\mathcal P}(d\tilde{\mathcal P})^4] =
32\,\tr\left[ (d\phi^\da \cdot d\phi)(d\phi^\da\cdot
d\phi)\right]+\op(\phi^6)$, where we suppress the wedge product
for notational convenience.

Therefore, after carrying out this procedure (and trivially
rescaling the coupling as $g \to g/8$), we obtain an alternative
form of Eq.~\ref{3dboundary} as a local (2+1)$d$ action in terms
of unconstrained boson fields. The field $\phi$ is a $n \times
(N-n)$ matrix, it has exactly the same number of degrees of
freedom as the target manifold $\mathcal{M}$, thus it does not
have any continuous gauge freedom. The Lagrangian density takes
the form
\begin{align}\label{eq:boundary action}
&\la = \la_{\text{NLSM}}+\la_{\text{WZW}}.
\end{align}
After rescaling $\phi \rightarrow \sqrt{g}\phi$, we find the
Euclidean Lagrangian density $\la_{\text{NLSM}}$ \beqn &&
\la_{\text{NLSM}} = \tr \left(\pa_\mu\phi^\da \cdot \pa_\mu\phi
\right) \cr\cr &+& \fourth g \tr \left[ \left( \pa_\mu\phi^\da
\cdot \phi +\phi^\da \cdot \pa_\mu\phi\right)^2 \right] \cr\cr &+&
\fourth g^\prime \tr \left[ \left(\pa_\mu\phi^\da \cdot \phi -
\phi^\da \cdot \pa_\mu\phi \right)^2\right] \cr\cr &+& \fourth g^2
\tr \left[ 2(\phi^\da \cdot \phi)(\pa_\mu\phi^\da \cdot
\phi)(\phi^\da \cdot \pa_\mu\phi) \right] \cr\cr &+& \fourth g^2
\tr \left[ (\phi^\da \cdot \phi)(\pa_\mu\phi^\da \cdot
\phi)(\pa_\mu\phi^\da \cdot \phi)\right] \cr\cr &+& \fourth g^2
\tr \left[(\phi^\da \cdot \phi) (\phi^\da \cdot
\pa_\mu\phi)(\phi^\da \cdot \pa_\mu\phi)\right] \cr\cr &+& \op(g^3
\phi^8). \label{lag}\eeqn The initial value of $g^\prime$ equals
to $g$, but under renormalization group flow it will be an
independent parameter from $g$. If we add more symmetry-allowed
terms in the original theory, they will only lead to obviously
irrelevant perturbations in the Lagrangian expanded in terms of
$\phi$.

After integrating over the $u$ direction in Eq.(\ref{3dboundary}),
the WZW term now reads \beqn \label{wzw} \la_{\text{WZW}}=
i\frac{k g^2}{4\pi} \e^{\mu\nu\rho} \tr \left[ \left(\phi^\da
\cdot\pa_\mu \phi \right) \left( \pa_\nu\phi^\da \cdot\pa_\rho\phi
\right)\right] - \cr\cr i \frac{k}{4\pi} g^3 \e^{\mu\nu\rho}
\frac{1}{3}\tr \left[ \left(\partial_\mu \phi^\da \cdot \phi
\right) \left(\partial_\nu \phi^\da \cdot \phi \right)
\left(\partial_\rho \phi^\da \cdot \phi \right) + h.c. \right]
\cr\cr + \op(k g^4 \phi^8)\;. \eeqn
%The Feynman rules for the propagator and ordinary 4-point vertex
%are standard.

It is convenient to adopt a double-line notation for the Feynman
diagrams, where a solid line represents $I = n+1, \cdots N$, and a
dashed line represents $\alpha = 1, \cdots n.$ We first compute
the ordinary RG equation in the large-$N$ limit without the WZW
term. We will calculate the beta function with $k=0$ in arbitrary
dimension $d$ and insert the physical value $d=3$. In terms of the
dimensionless coupling $\tilde g = \Ld^{d-2} g$ and $\tilde
g^\prime = \Ld^{d-2} g^\prime$ ($\Ld \sim 1/l$ is the ultraviolet
momentum cut-off), the beta functions in the large-$N$ limit for
the ordinary NLSM (with $k=0$) are \beqn \label{eq:ordinary beta
function} \beta(\tilde g)_0 = \frac{d\tilde g}{d \ln l } &=& -
(d-2) \tilde g + \frac{N}{2\pi^2} \tilde g^2, \cr\cr \beta(\tilde
g^\prime)_0 = \frac{d\tilde g^\prime}{d \ln l } &=& - (d-2)\tilde
g^\prime + \frac{N}{d \pi^2} \tilde g^{\prime 2}, \eeqn in our
current case $d = 3$. As long as $n \sim N^{A}$ with $A<1$, in the
large-$N$ limit we only need to keep these terms in the beta
functions. Eq.~\ref{eq:ordinary beta function} has several fixed
points. If we start with the physical parameter $\tilde g(\Lambda)
= \tilde g^\prime (\Lambda)$ as the tuning parameter at the
beginning of the RG flow, then increasing $\tilde g$ will lead to
a quantum phase transition controlled by the fixed point \beqn
\tilde g_\ast = \frac{2\pi^2}{N}, \ \ \ \tilde g^\prime_\ast = 0,
\label{fixed points} \eeqn and the critical exponent $\nu = 1$.
The location of the critical point, and the critical exponent is
consistent with the well-known result of the CP$^{N-1}$ model in
the large$-N$ limit~\cite{halperincp,kaulsachdev}.

\section{Stable fixed point in the quantum disordered phase}

\begin{figure}
\includegraphics[width=0.55\columnwidth]{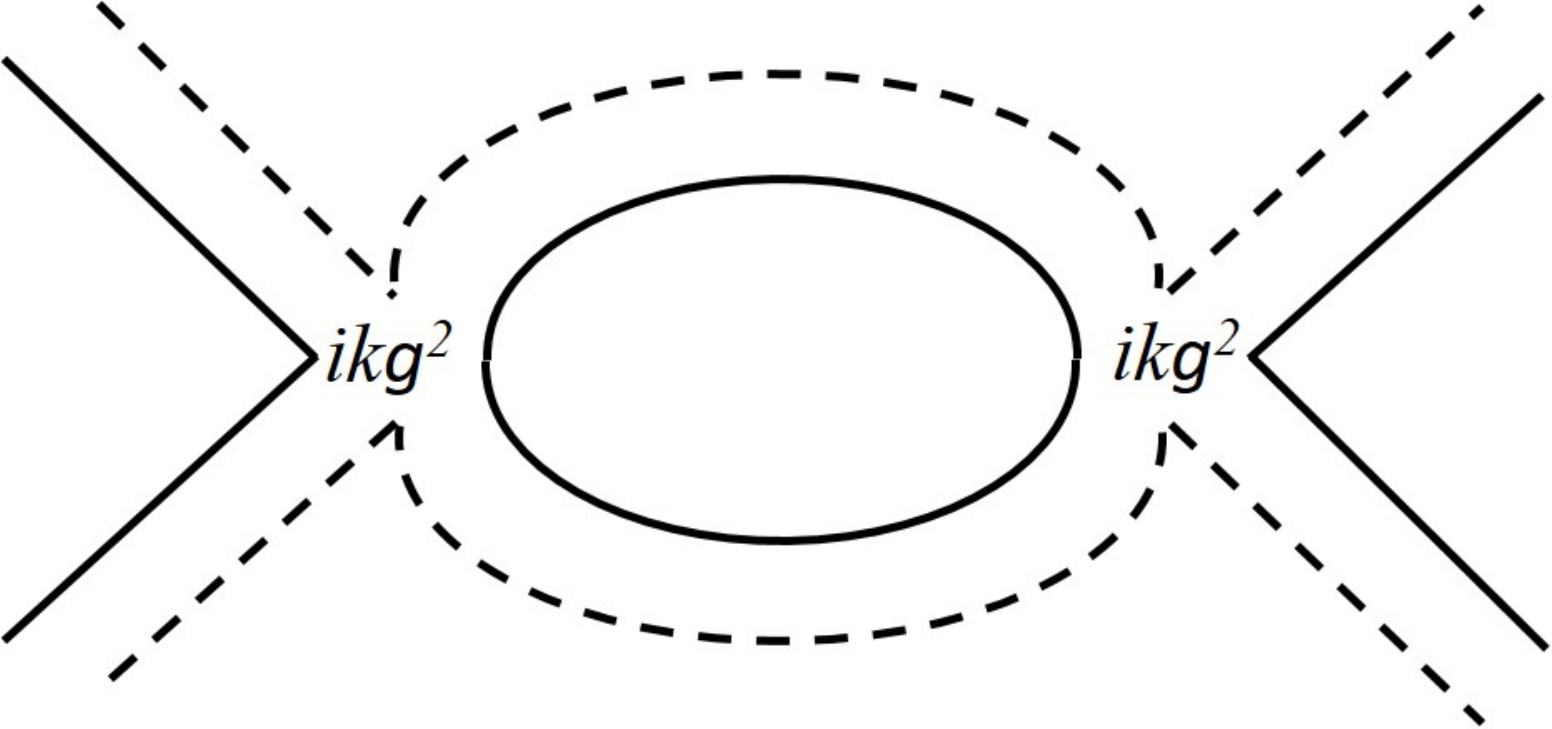}
\caption{One-loop diagram which involves two WZW terms in
Eq.~\ref{wzw}. The numerator of the WZW vertex is completely
antisymmetric in momenta, so this diagram does not renormalize $g$
or $g^\prime$.}\label{1loop}
\end{figure}

\begin{figure}
\includegraphics[width=0.55\columnwidth]{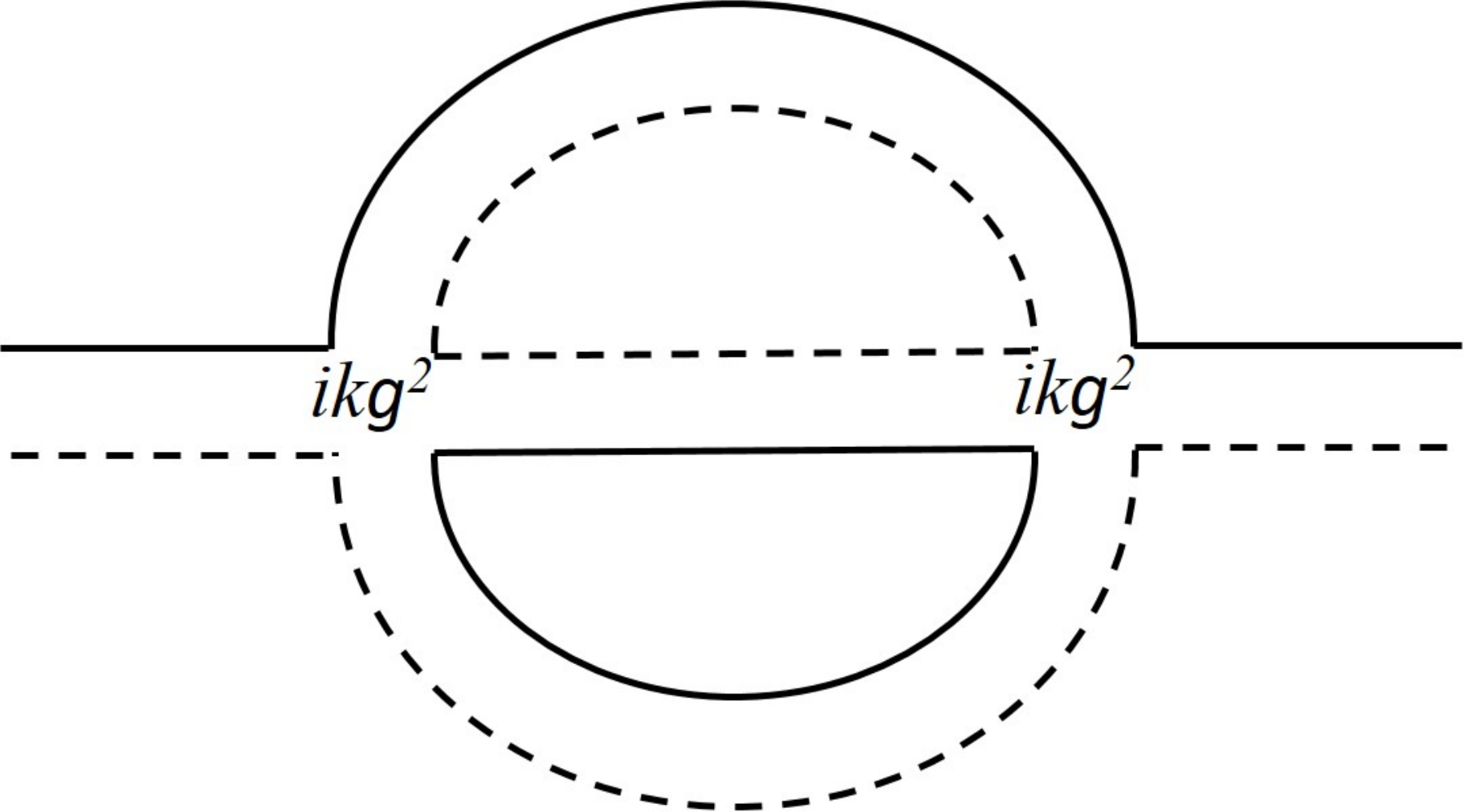}
\caption{Two-loop wave function renormalization.}\label{2loop}
\end{figure}

%\subsection{Infinite diagram difficulty}

Now let us compute the beta functions with the WZW term. Naively
one would expect that the leading order contribution from the WZW
term to the beta functions is the one-loop diagram
Fig.~\ref{1loop}. But because the numerator of the WZW vertex is
completely antisymmetric in momenta, this diagram does not
renormalize the coupling constants $g$ and $g^\prime$.
Fig.~\ref{2loop} is a two-loop planar wave function
renormalization diagram that renormalizes $g$ and $g^\prime$. This
diagram leads to the following corrections to the beta functions:
\beqn \label{eq:2loop diagram} \beta(\tilde g) &=& \beta(\tilde
g)_0 - c k^2 \tilde g^5 N n \frac{1}{(4\pi)^2} + \cdots  \cr\cr
\beta(\tilde g^\prime) &=& \beta(\tilde g^\prime)_0 - c k^2 \tilde
g^\prime \tilde g^4 N n \frac{1}{(4\pi)^2} + \cdots\eeqn In this
equation $c$ is a positive number whose exact value is
unimportant, because we are going to treat $k^2$ as a tuning
parameter.

\begin{figure}
\includegraphics[width=0.6\columnwidth]{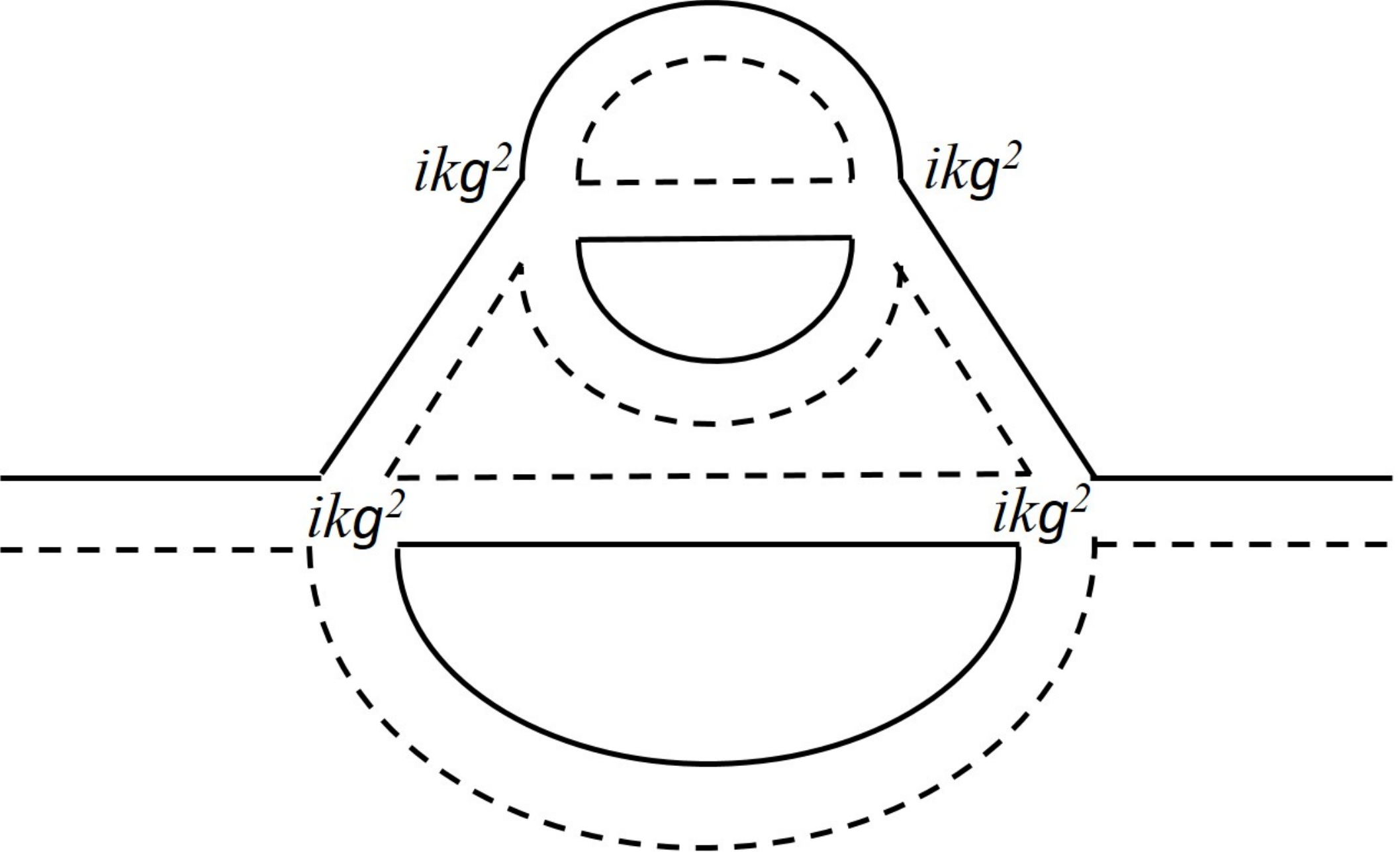}
\caption{A four-loop diagram with four WZW terms. When $k \leq
N^{3/2}$, in the simultaneous limit of large $N$ and large $k$,
this diagram contributes to the RG equation at least at the same
order as the two-loop diagram in Fig.~\ref{2loop} around the ``new
stable fixed point" in the quantum disordered phase, so do
infinite number of higher loop diagrams.} \label{fig:4loop}
\end{figure}

Our goal is to look for a stable fixed point which corresponds to
a stable $(2+1)d$ CFT in the quantum disordered phase. The
negative sign of the $k^2$ term in Eq.~\ref{eq:2loop diagram}
suggests that this is possible. However, to make a confident
conclusion, we need to choose certain adequate scaling between $k$
and $N$: $k\sim N^B$. If for instance $0 < B \leq 3/2$, then the
$k^2$ terms in Eq.~\ref{eq:2loop diagram} indeed lead to a new
{\it stable} fixed point in the quantum disordered phase at
$\tilde g_\ast \sim k^{-2/3} \geq 1/N$ and $\tilde g^\prime_\ast =
0$. But around this ``new fixed point", infinite number of higher
loop diagrams would become nonperturbative. For example, let us
examine the four-loop WZW contribution, which is shown in
Fig.~\ref{fig:4loop}. This diagram has seven internal propagators,
four WZW vertices, two closed solid loops, and two closed dashed
loops. Therefore this diagram contributes a term $\sim g^9 k^4 n^2
N^2$ to the beta function. Then when $B \leq 3/2$, this four-loop
diagram (and infinite number of higher loop diagrams) also
contributes at least at the same order as the $k^2$ terms in
Eq.~\ref{eq:2loop diagram}, around the ``new fixed point" $\tilde
g_\ast \sim k^{-2/3}$.

But on the other hand, if $B > 3/2$, then the $k^2$ term in
Eq.~\ref{eq:2loop diagram} would be too large and make the entire
RG equations flow to $\tilde g = \tilde g^\prime = 0$. We stress
that these difficulties only occur with the presence of the WZW
term. Without the WZW term, this theory does have a simple
large-$N$ limit.

%\subsection{Stable CFT fixed point with $n=1$}

In order to find a controlled calculation and to identify the
stable fixed point in the quantum disordered phase with
confidence, we need to find another small parameter to expand
with. As we mentioned before we cannot rely on the ordinary $2 +
\epsilon$ expansion in our case. In this section we propose a
possible solution to this difficulty in our current context by
introducing a different $\epsilon-$generalization of our model.

We first test our approach with $n=1$ ($\mathcal{M} =
\mathrm{CP}^{N-1} $).
%and we will present {\it two}
%$\epsilon-$generalizations that differ technically.
%In our first method,
We generalize the original action Eq.~\ref{eq:boundary action} as
following: \beqn && \la_{\text{NLSM}} = \pa_\mu\phi^\da \cdot
\pa_\mu\phi \cr\cr &+& \fourth g \left( \pa_\mu\phi^\da \cdot
|\bar{\partial}|^{\frac{\epsilon-1}{2}} \phi + \phi^\da \cdot
|\bar{\partial}|^{\frac{\epsilon-1}{2}} \pa_\mu\phi \right)^2
\cr\cr &+& \fourth g^\prime \left(\phi^\dagger \cdot
|\bar{\partial}|^{\frac{\epsilon-1}{2}} \partial_\mu \phi -
\partial_\mu \phi^\dagger \cdot
|\bar{\partial}|^{\frac{\epsilon-1}{2}} \phi \right)^2 \cr\cr &+&
\fourth g^2 \left(\phi^\da \cdot |\bar{\partial}|^{\epsilon-1}
\phi \right) \left(\pa_\mu\phi^\da \cdot
|\bar{\partial}|^{\frac{\epsilon-1}{2}} \phi + \phi^\da \cdot
|\bar{\partial}|^{\frac{\epsilon-1}{2}} \pa_\mu\phi \right)^2
\cr\cr &+& \op(g^3 \phi^8). \label{newaction}\eeqn Here the
notation $|\bar{\partial}|$ is most manifest in the momentum
space: $A^\dagger |\bar{\partial}| B$ in the momentum space
corresponds to $A^\dagger(\vec{p})|\frac{1}{2}(\vec{p} + \vec{q})|
B(\vec{q})$. This nonanalytic generalization can be made
systematically to all higher order expansion of the Lagrangian: a
singular momentum dependence
$|\bar{\partial}|^{\frac{\epsilon-1}{2}}$ is inserted in
$\phi^\dagger \cdot
\partial_\mu \phi$ and $\phi \cdot
\partial_\mu \phi^\dagger$, and $|\bar{\partial}|^{\epsilon - 1}$ is inserted in
$\phi^\dagger \cdot \phi$. At least in the large$-N$ limit, it can
be shown that all the relevant renormalizations to this Lagrangian
can still be absorbed into the RG flow of $g$ and $g^\prime$.

The nonanalytic generalization of a local field theory dated back
to studies on spin systems with long range
interactions~\cite{fisher1972}, and the study of the Gross-Neveu
model~\cite{GNrenormalizable}. Later a generalization of the
regular $p^2$ kinetic term to $|p|^{1+\epsilon}$ was used as a
controlled calculation method for $2d$ Fermi surface coupled with
a bosonic field~\cite{senthilfermi,chetanfermi1,chetanfermi2},
which without the nonanalytic generalization also suffers from the
infinite diagram difficulty in the large-$N$ limit~\cite{sslee}.
The advantage of the nonanalytic generalization is that, now the
scaling dimension of $g$ and $g^\prime$ at weak interacting limit
becomes $-\epsilon$, and we can treat $\epsilon$ as another small
parameter to organize all the Feynman diagrams.

The WZW term is now generalized to \beqn \label{newwzw}
\la_{\text{WZW}}= i\frac{k g^2}{4\pi} \e^{\mu\nu\rho} \left(
\phi^\da \cdot |\bar{\partial}|^{\epsilon-1} \pa_\mu \phi \right)
\left(\pa_\nu\phi^\da \cdot |\bar{\partial}|^{\epsilon-1}
\pa_\rho\phi \right). \eeqn When $n=1$ there is no higher order
terms in the WZW term, which significantly simplifies the
analysis. When $\epsilon = 1$ this action returns to its original
form Eq.~\ref{eq:boundary action}.

This generalization keeps many of the basic properties of the
original WZW term:

{\it 1)} this term Eq.~\ref{newwzw} is always purely imaginary;

{\it 2)} like the WZW term, the parameter $k$ is always marginal
for arbitrary $\epsilon$, which is guaranteed by the nonanalytic
momentum dependence inserted in the generalized WZW term;

{\it 3)} the two $\phi$ ($\phi^\dagger$) fields in
Eq.~\ref{newwzw} are equivalent to each other.

\begin{figure}
\includegraphics[width=0.9\columnwidth]{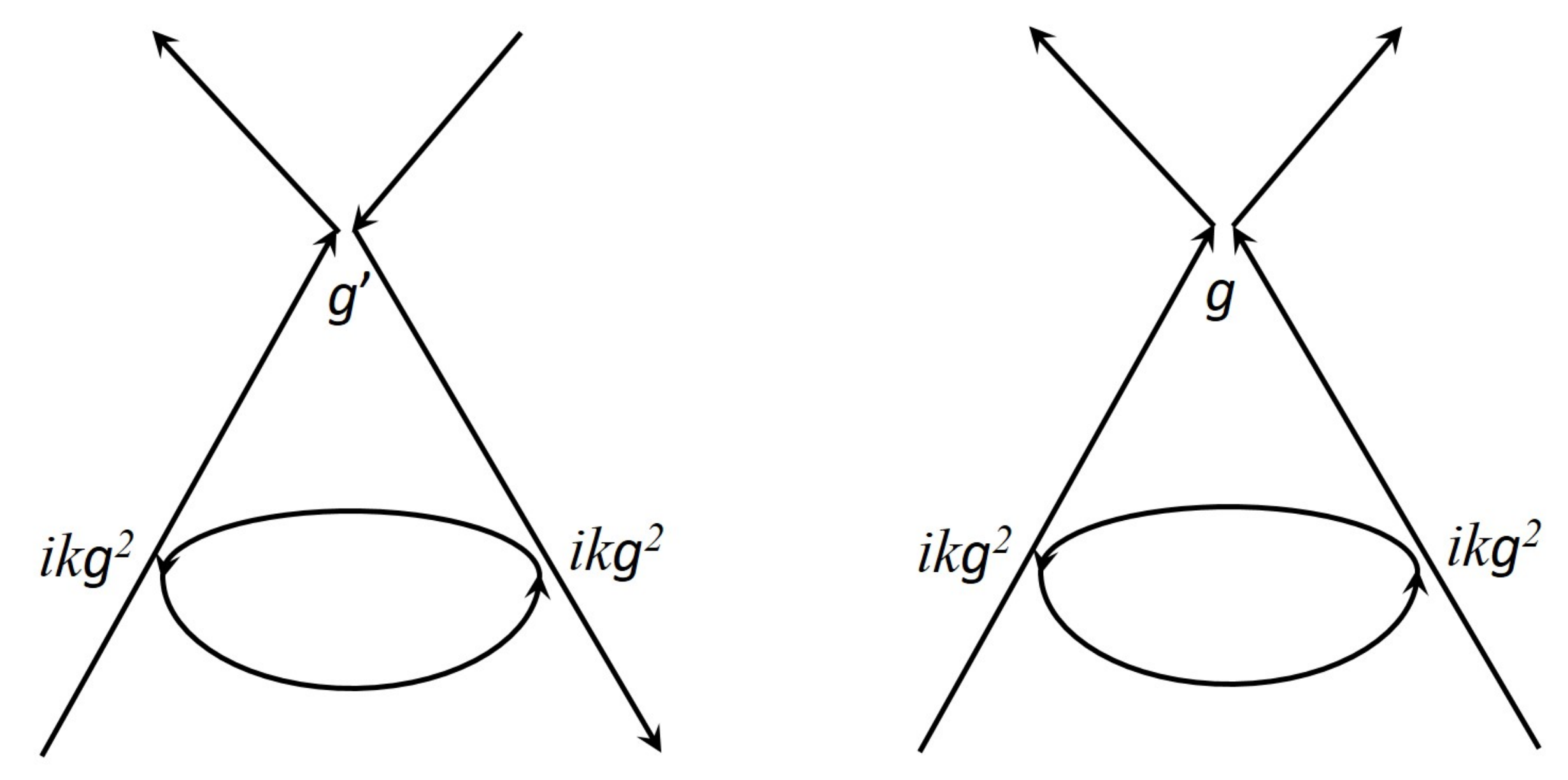}
\caption{The vertex corrections from the WZW terms, which generate
irrelevant interactions under RG with our nonanalytic
$\epsilon-$generalization.} \label{vertex}
\end{figure}

With large-$N$ and leading order in $\epsilon$, the RG equations
of $\tilde{g}$ and $\tilde g^\prime$ read (here we redefine
$\tilde g=\Lambda^\epsilon g$ and $\tilde
g^\prime=\Lambda^\epsilon g^\prime$ to make them dimensionless)
\beqn \label{newRG} \frac{d\tilde g}{d \ln l } &=& \beta(\tilde
g)_0^{(\epsilon)} - c k^2 \tilde g^5 N \frac{1}{(4\pi)^2}, \cr\cr
\frac{d\tilde g^\prime}{d \ln l} &=& \beta(\tilde
g^\prime)_0^{(\epsilon)} - c k^2 \tilde g^\prime \tilde g^4 N
\frac{1}{(4 \pi)^2}. \eeqn $\beta(\tilde g)_0^{(\epsilon)}$ and
$\beta(\tilde g^\prime)_0^{(\epsilon)}$ are simply $\beta(\tilde
g)_0$ and $\beta(\tilde g^\prime)_0$ with the first term replaced
by $ - \epsilon \tilde g$ and $ - \epsilon \tilde g^\prime$. The
wave function renormalization Fig.~\ref{2loop} is the only diagram
that contributes to the last terms in Eq.~\ref{newRG} in the
large$-N$ limit. Vertex corrections in Fig.~\ref{vertex} will not
contribute here because under RG flow it generates an $\phi^4$
term with analytic momentum dependence, which is less relevant
compared with the terms in Eq.~\ref{newaction}. The absence of
vertex corrections here is similar to the absence of boson field
wave function renormalization discussed in
Ref.~\onlinecite{chetanfermi1}, basically because a nonanalytic
momentum dependence cannot be generated by integrating out high
momentum degrees of freedom in RG. This absence of vertex
correction to terms with nonanalytic momentum dependence was also
discussed in Ref.~\onlinecite{freybalents,xumuellersachdev}.

Now we need to take $k^2 \sim (N/\epsilon)^3$ to keep all the
terms in these equations at the same order, and we expect that the
fixed points of these beta functions will be around $\tilde g \sim
\epsilon/N$. With small enough $\epsilon$, the terms we keep in
Eq.~\ref{newRG} will be dominant compared with all higher loop
diagrams.

The value of constant $c$ is computed at $\epsilon =0$: with
large-$N$, large-$k$ and $\epsilon = 0$, the wave function
renormalization in Fig.~\ref{2loop} will lead to the following
correction to the coupling constant $g$: \beqn \delta \tilde g &=&
- 8 \tilde g^5 N \left( \frac{k}{4\pi} \right)^2 \int
\frac{d^3p}{(2\pi)^3} \frac{d^3q}{(2\pi)^3} \cr\cr &\times&
\frac{1}{3} \frac{p^2 q^2 - (\vec{p} \cdot \vec{q})^2 }{p^2 q^2
|\vec{p} + \vec{q}|^{2} |\vec{p} - \vec{q}|^{4}} \times 16 \cr\cr
&\sim& - \frac{1}{3 \pi^2} k^2 \tilde g^5 N \frac{1}{(4\pi)^2}
\log \left( \frac{\Lambda}{\Lambda ^\prime} \right), \eeqn where
$\Lambda$ and $\Lambda^\prime$ are the ultraviolet cut-off and
rescaled cut-off. Thus $c = 1/(3 \pi^2)$. The value of $c$
evaluated at $\epsilon = 0$ depends on the exact form of the
$\epsilon$ generalization of the WZW term.
%For example, if we replace $|\bar{\partial}|^{\epsilon -
%1}$ by $|2\bar{\partial}|^{\epsilon - 1}$, $c$ will become $c =
%1/(48 \pi^2)$.

%\begin{figure}
%\includegraphics[width=0.55\columnwidth]{2loop_vt.pdf}
%\caption{Two-loop vertex correction to $g^\prime$.}\label{2loop_v}
%\end{figure}

%\begin{figure}
%\includegraphics[width=\columnwidth]{RG2.pdf}
%\caption{The RG flow diagram for the RG equations in
%Eq.~\ref{eq:2loop diagram}, truncated at the two-loop planar
%diagrams Fig.~\ref{2loop},\ref{2loop_v},\ref{fish}. We chose
%parameters $N = 200$, $ck^2n \sim 85$.} \label{RG}
%\end{figure}

We take $k^2 = G^3 (N/\epsilon)^3$ with small coefficient $G$.
Eq.~\ref{newRG} generates several fixed points. If we start with
the physical parameters $\tilde g(\Lambda) = \tilde
g^\prime(\Lambda)$ at the beginning of the RG, the flow of the
parameters is controlled by two of these fixed points. The first
fixed point is the order-disorder quantum phase transition located
at \beqn \tilde g_\ast \sim (2 \pi^2 + 2 \pi^8 c G^3 +
\mathcal{O}(G^6)) \frac{\epsilon}{N}, \ \ \ \tilde g^\prime_\ast =
0 \label{qcp2}\eeqn and the critical exponent $1/\nu$ is \beqn
\frac{1}{\nu} = \epsilon (1 - 3 c G^3\pi^6 + \mathcal{O}(G^6)).
\eeqn If we extrapolate $\epsilon$ to $1$, $\nu$ will be greater
than 1, which can already be expected from the negative sign of
the $k^2$ term in the beta functions. This is qualitatively
different from the critical exponent without the WZW term. For
instance it is well-known that the $(2 + 1)d$ CP$^{N-1}$ model has
$\nu < 1$ with $1/N$ correction taken into
account~\cite{kaulsachdev}.

\begin{figure}
\includegraphics[width=\columnwidth]{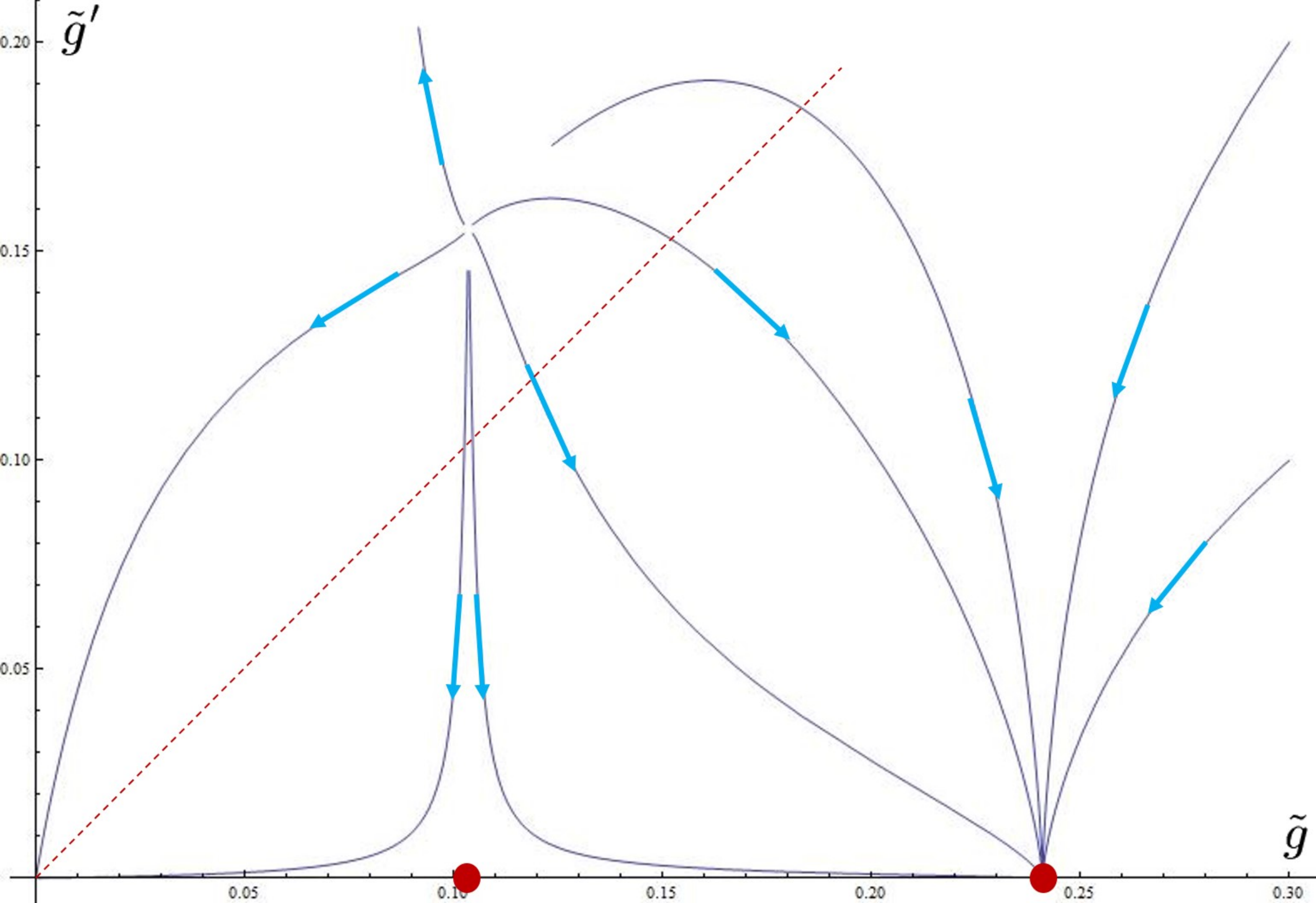}
\caption{The RG flow diagram for the RG equations in
Eq.~\ref{newRG}. We chose parameters $\epsilon = 0.05$ $N = 10$,
$ck^2n \sim 340$. The dashed line corresponds to the physical
values of the tuning parameter $\tilde g = \tilde g^\prime$ at the
beginning of the RG flow. The RG flow is controlled by two fixed
points, one is the order-disorder transition, the other is a
stable fixed point in the quantum disordered phase.} \label{RG}
\end{figure}

Most importantly, there is a stable fixed point in the quantum
disordered phase: \beqn \tilde g_\ast \sim \left( \frac{1}{G}
\frac{2}{c^{1/3}} - \frac{2\pi^2}{3} + \mathcal{O}(G) \right)
\frac{\epsilon}{N}, \ \ \ \tilde g_\ast^\prime = 0.
\label{fp}\eeqn We need $G$ small enough to guarantee that the
coupling constant in Eq.~\ref{fp} is larger than the one in
Eq.~\ref{qcp2}, $i.e.$ the system is in a quantum disordered
phase. In the vicinity of this new stable fixed points, the beta
functions give the scaling dimension of two irrelevant
perturbations: \beqn \Delta_1 &=& \epsilon \left( - \frac{1}{G}
\frac{3}{c^{1/3} \pi^2} + 5 + \mathcal{O}(G) \right), \cr \cr
\Delta_2 &=& \epsilon \left( - \frac{1}{G} \frac{1}{c^{1/3} \pi^2}
+ \frac{1}{3} + \mathcal{O}(G) \right). \eeqn Both scaling
dimensions are negative with small enough $G$. The RG flow diagram
for the RG equations with parameters $\epsilon = 0.05$ $N = 10$,
$ck^2n \sim 340$ is plotted in Fig.~\ref{RG}.

%\subsection{Case with $n > 1$}

\begin{figure}
\includegraphics[width=0.55\columnwidth]{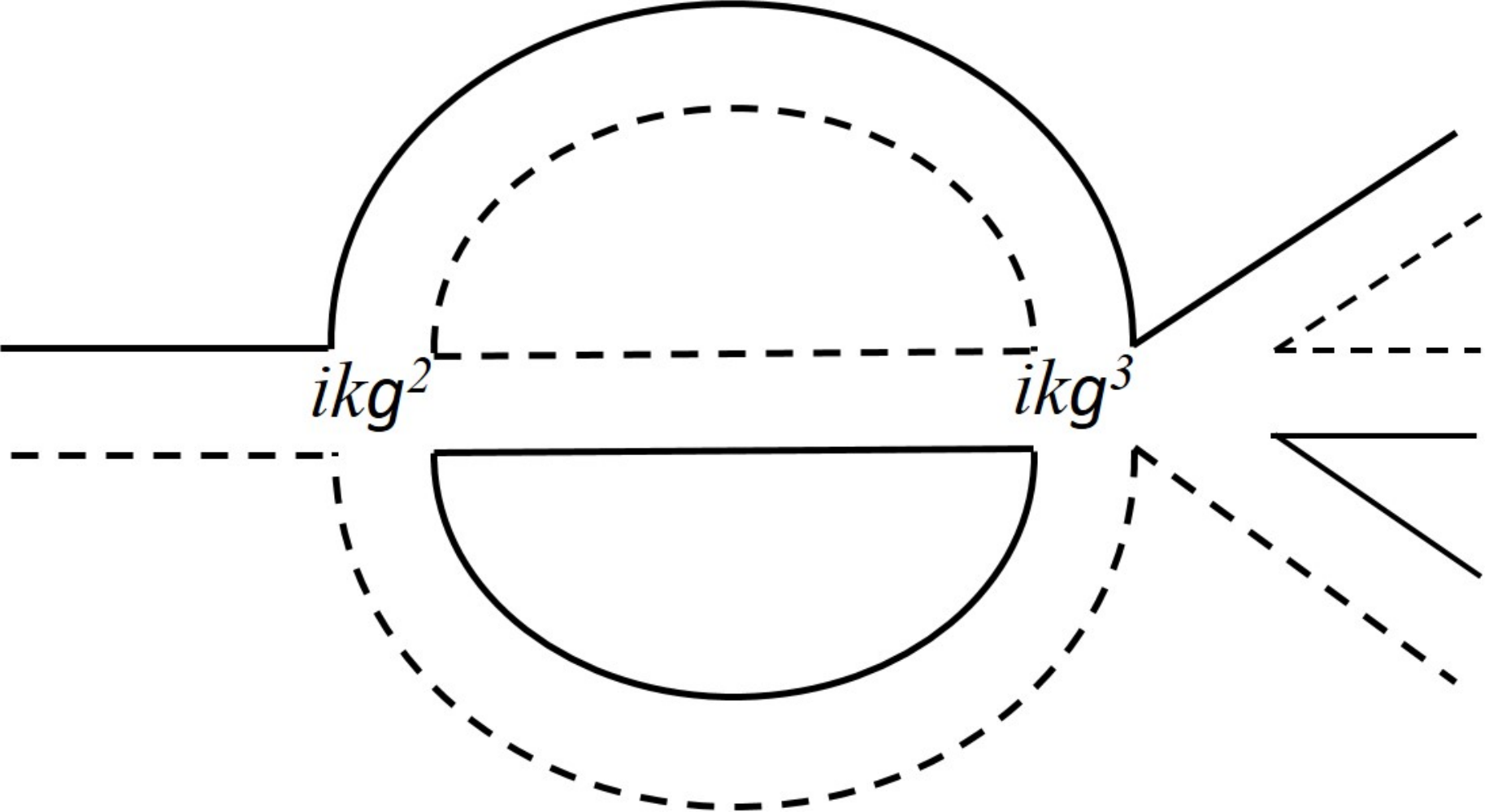}
\caption{A two loop diagram that is a mixture between the $\phi^4$
and $\phi^6$ terms in the WZW term for $n>1$.} \label{fish}
\end{figure}

In order to carry out the calculation for $n > 1$, we need to
include higher order terms in the expansion of the WZW term. We
also need to generalize the $\mathcal{O}(\phi^6)$ order in the WZW
term to a nonanalytic form. There are certainly more than one
possible $\epsilon-$generations, as an example, we choose the
following form for the $\phi^6$ term in the momentum space: \beqn
\la_{\text{WZW}}(\phi^6) =
 - \sum_{\vec{w}, \vec{l}, \vec{p},
\vec{q},\vec{t},\vec{s}} \delta( \vec{w} + \vec{p} + \vec{s} -
\vec{l} - \vec{q} - \vec{r}) \cr\cr \times \frac{k g^3}{4\pi}
\frac{1}{3} \e^{\mu\nu\rho} l_\mu q_\nu t_\rho | \vec{l} +
\vec{p}|^{\epsilon-1} |\vec{q} + \vec{s}|^{\epsilon-1}|\vec{t} +
\vec{w}|^{\epsilon-1} \cr\cr \times \mathrm{tr} \left(
\phi^\dagger (\vec{l}) \cdot  \phi (\vec{w})
 \ \phi^\dagger (\vec{q}) \cdot \phi(\vec{p})  \
\phi^\dagger (\vec{t}) \cdot  \phi (\vec{s}) - h.c. \right). \eeqn
This generalization still keeps the basic properties of the WZW
term that we need to carry out the calculations, and when
$\epsilon = 1$ it returns to the original form of the WZW term.
This $\phi^6$ term so designed only generates irrelevant terms in
the large$-N$ limit and leading order $\epsilon$ expansion. For
example, Fig.~\ref{fish} is a leading order diagram in terms of
large$-N$ and $\epsilon-$expansion counting, but it only generates
an irrelevant analytic term to the Lagrangian.

%However, the $\epsilon-$expansion for the case of $n > 1$ is
%technically more complicated due to the fact that, for each higher
%order of $\epsilon$ expansion, we need to involve a higher order
%term in the expansion of the WZW term. And it is likely that we
%need to individually design a $\epsilon-$generalization for each
%order, so our method is not ideally convenient for $n>1$.

\section{Discussions}

%Technically there are other options of carrying out the RG
%calculation. For instance, what is often used for NLSM is the
%background field method, where the constrained field is decomposed
%into a slow moving background configuration and a fast moving
%fluctuation. In simpler models this method does lead to the same
%beta function as the approach we are using in this work. However,
%the Grassmannian manifold we discuss in this paper is slightly too
%complicated for a background field method.

In this work we did our best to search for a controlled study of
stable interacting conformal field theories at the boundary of
$(3+1)d$ SPT states. We performed calculation in the large$-N$
limit and leading order $\epsilon-$expansion, and the desired
stable fixed point is indeed found in the quantum disordered
phase. But we have not proved that higher order expansions will
not generate more relevant terms in the Lagrangian.

%For $n=1$, our calculation is reliable with small $\epsilon$ and
%certain scaling limit between $N$ and $k$. For $n>1$, our method
%is less convenient and has room for improvement. Just like all the
%other $\epsilon$ expansions, eventually we will need to
%extrapolate the calculation to $\epsilon = 1$. Whether our
%expansion of $\epsilon$ converges with $\epsilon = 1$ depends on
%the radius of convergence of this expansion.

Besides exploring the exotic boundary states of $(3+1)d$ bosonic
SPT phases, another motivation of this work was the ``hierarchy
problem" in high energy physics: why the Higgs boson is so much
lighter than the Planck mass? Compared with the Planck mass, the
Higgs boson, which is a space-time scalar, is almost massless.
Gauge bosons, which can emerge very naturally in condensed matter
systems~\cite{wenphoton,sondhiphoton,hermelephoton}, indeed have
zero mass. But a space-time scalar boson, unless it is a Goldstone
mode, usually acquires a mass that is comparable with the
ultraviolet cut-off without fine-tuning to a critical point. At
least this is the case for space-time dimensions higher than
$(1+1)d$ (in $(1+1)d$ space-time scalar bosons can easily form a
conformal field theory). Indeed, the little Higgs theory
hypothesizes that the Higgs boson itself is a pseudo Goldstone
boson~\cite{littlehiggs,littlehiggs1,littlehiggs2,littlehiggs3},
which explains its small mass. The result of our current work
suggests another possible route to address the hierarchy problem:
the Higgs boson could be rendered massless due to a topological
WZW term, even if the system is in a quantum disordered phase,
$i.e.$ there is no (pseudo) spontaneous symmetry breaking. But, in
order to show this explicitly, one needs to first embed the Higgs
boson into a larger target manifold $\mathcal{M}$ which permits a
WZW term, and perform a controlled RG calculation in
$(3+1)d$~\footnote{$\pi_5[SU(N)] = \mathbb{Z}$ for $N>2$, thus a
matrix model whose target manifold is $SU(N)$ could have a WZW
term in $(3+1)d$. But $SU(N)$ matrix model does not have a
controlled large$-N$ limit even without the WZW term.}. We will
leave this direction to future study.

At the purely technical level, although the WZW term can be
formally rewritten as a Chern-Simons term,
%one may attempt to study the theory
%discussed in this paper by interpreting the gauge field
we cannot treat the gauge field $a_\mu$ (Eq.~\ref{gauge}) in the
path integral as if it were an independent degree of freedom with
a Chern-Simons term. For example, when $N = 2n = 2$, the
topological term becomes the quantized Hopf term if written in
terms of $\varphi^i$, while the Chern-Simons action of a U(1)
gauge field is in general not quantized. The WZW term can only be
interpreted as the Chern-Simons term if Eq.~\ref{gauge} holds
rigorously. However, if a Chern-Simons term of $a_\mu$ is already
included in the action, the equation of motion of the gauge field
is no longer given by Eq.~\ref{gauge}. In the standard path
integral formalism of the CP$^{N-1}$ model, the gauge field
$a_\mu$ is introduced as an auxiliary field through the
Hubbard-Stratonovich transformation. Thus one should introduce one
more vector field $b_\mu$ through the Hubbard-Stratonovich
transformation on the WZW term: $\sim ik \varepsilon
\varphi^\dagger \cdot \partial \varphi
\partial b + i k \varepsilon b
\partial b $ (indices and unimportant factors are omitted in this equation).
Integrating out $b_\mu$ will regenerate the WZW term, for the
simplest case $n=1$. For $n > 1$ this method gets more
complicated.

%The simplest
%example is the $(2+1)d$ O(3) NLSM with a Hopf-$\Theta$ term, due
%to the fact $\pi_3[S^2] = \mathbb{Z}$. The Hopf term can be
%naively rewritten as a U(1) CS term with level $k = \Theta/\pi$.
%However, physically we understand that when $\Theta = 2\pi k$, the
%quantum disordered phase

Bi, Rasmussen and Xu are supported by the David and Lucile Packard
Foundation and NSF Grant No. DMR-1151208. BenTov is supported by
the Simons Foundation Agency with award number: 376205. The
authors thank Leon Balents and Andreas Ludwig for helpful
discussions.

\bibliography{NK1}

\begin{thebibliography}{42}
\expandafter\ifx\csname natexlab\endcsname\relax\def\natexlab#1{#1}\fi
\expandafter\ifx\csname bibnamefont\endcsname\relax
  \def\bibnamefont#1{#1}\fi
\expandafter\ifx\csname bibfnamefont\endcsname\relax
  \def\bibfnamefont#1{#1}\fi
\expandafter\ifx\csname citenamefont\endcsname\relax
  \def\citenamefont#1{#1}\fi
\expandafter\ifx\csname url\endcsname\relax
  \def\url#1{\texttt{#1}}\fi
\expandafter\ifx\csname urlprefix\endcsname\relax\def\urlprefix{URL }\fi
\providecommand{\bibinfo}[2]{#2}
\providecommand{\eprint}[2][]{\url{#2}}

\bibitem[{\citenamefont{Chen et~al.}(2013)\citenamefont{Chen, Gu, Liu, and
  Wen}}]{wenspt}
\bibinfo{author}{\bibfnamefont{X.}~\bibnamefont{Chen}},
  \bibinfo{author}{\bibfnamefont{Z.-C.} \bibnamefont{Gu}},
  \bibinfo{author}{\bibfnamefont{Z.-X.} \bibnamefont{Liu}}, \bibnamefont{and}
  \bibinfo{author}{\bibfnamefont{X.-G.} \bibnamefont{Wen}},
  \bibinfo{journal}{Phys. Rev. B} \textbf{\bibinfo{volume}{87}},
  \bibinfo{pages}{155114} (\bibinfo{year}{2013}).

\bibitem[{\citenamefont{Chen et~al.}(2012)\citenamefont{Chen, Gu, Liu, and
  Wen}}]{wenspt2}
\bibinfo{author}{\bibfnamefont{X.}~\bibnamefont{Chen}},
  \bibinfo{author}{\bibfnamefont{Z.-C.} \bibnamefont{Gu}},
  \bibinfo{author}{\bibfnamefont{Z.-X.} \bibnamefont{Liu}}, \bibnamefont{and}
  \bibinfo{author}{\bibfnamefont{X.-G.} \bibnamefont{Wen}},
  \bibinfo{journal}{Science} \textbf{\bibinfo{volume}{338}},
  \bibinfo{pages}{1604} (\bibinfo{year}{2012}).

\bibitem[{\citenamefont{Lu and Vishwanath}(2012)}]{luashvin}
\bibinfo{author}{\bibfnamefont{Y.-M.} \bibnamefont{Lu}} \bibnamefont{and}
  \bibinfo{author}{\bibfnamefont{A.}~\bibnamefont{Vishwanath}},
  \bibinfo{journal}{Phys. Rev. B} \textbf{\bibinfo{volume}{86}},
  \bibinfo{pages}{125119} (\bibinfo{year}{2012}).

\bibitem[{\citenamefont{Fidkowski et~al.}(2013)\citenamefont{Fidkowski, Chen,
  and Vishwanath}}]{TI_fidkowski1}
\bibinfo{author}{\bibfnamefont{L.}~\bibnamefont{Fidkowski}},
  \bibinfo{author}{\bibfnamefont{X.}~\bibnamefont{Chen}}, \bibnamefont{and}
  \bibinfo{author}{\bibfnamefont{A.}~\bibnamefont{Vishwanath}},
  \bibinfo{journal}{Phys. Rev. X} \textbf{\bibinfo{volume}{3}},
  \bibinfo{pages}{041016} (\bibinfo{year}{2013}).

\bibitem[{\citenamefont{Chen et~al.}(2014)\citenamefont{Chen, Fidkowski, and
  Vishwanath}}]{TI_fidkowski2}
\bibinfo{author}{\bibfnamefont{X.}~\bibnamefont{Chen}},
  \bibinfo{author}{\bibfnamefont{L.}~\bibnamefont{Fidkowski}},
  \bibnamefont{and}
  \bibinfo{author}{\bibfnamefont{A.}~\bibnamefont{Vishwanath}},
  \bibinfo{journal}{Phys. Rev. B} \textbf{\bibinfo{volume}{89}},
  \bibinfo{pages}{165132} (\bibinfo{year}{2014}).

\bibitem[{\citenamefont{Bonderson et~al.}(2013)\citenamefont{Bonderson, Nayak,
  and Qi}}]{TI_Qi}
\bibinfo{author}{\bibfnamefont{P.}~\bibnamefont{Bonderson}},
  \bibinfo{author}{\bibfnamefont{C.}~\bibnamefont{Nayak}}, \bibnamefont{and}
  \bibinfo{author}{\bibfnamefont{X.-L.} \bibnamefont{Qi}}, \bibinfo{journal}{J.
  Stat. Mech.} p. \bibinfo{pages}{P09016} (\bibinfo{year}{2013}).

\bibitem[{\citenamefont{Metlitski et~al.}(2013)\citenamefont{Metlitski, Kane,
  and Fisher}}]{TI_max}
\bibinfo{author}{\bibfnamefont{M.~A.} \bibnamefont{Metlitski}},
  \bibinfo{author}{\bibfnamefont{C.~L.} \bibnamefont{Kane}}, \bibnamefont{and}
  \bibinfo{author}{\bibfnamefont{M.~P.~A.} \bibnamefont{Fisher}},
  \bibinfo{journal}{arXiv:1306.3286}  (\bibinfo{year}{2013}).

\bibitem[{\citenamefont{Wang et~al.}(2013)\citenamefont{Wang, Potter, and
  Senthil}}]{TI_senthil}
\bibinfo{author}{\bibfnamefont{C.}~\bibnamefont{Wang}},
  \bibinfo{author}{\bibfnamefont{A.~C.} \bibnamefont{Potter}},
  \bibnamefont{and} \bibinfo{author}{\bibfnamefont{T.}~\bibnamefont{Senthil}},
  \bibinfo{journal}{Phys. Rev. B} \textbf{\bibinfo{volume}{88}},
  \bibinfo{pages}{115137} (\bibinfo{year}{2013}).

\bibitem[{\citenamefont{Vishwanath and Senthil}(2013)}]{senthilashvin}
\bibinfo{author}{\bibfnamefont{A.}~\bibnamefont{Vishwanath}} \bibnamefont{and}
  \bibinfo{author}{\bibfnamefont{T.}~\bibnamefont{Senthil}},
  \bibinfo{journal}{Phys. Rev. X} \textbf{\bibinfo{volume}{3}},
  \bibinfo{pages}{011016} (\bibinfo{year}{2013}).

\bibitem[{\citenamefont{Bi et~al.}(2015)\citenamefont{Bi, Rasmussen, and
  Xu}}]{xuclass}
\bibinfo{author}{\bibfnamefont{Z.}~\bibnamefont{Bi}},
  \bibinfo{author}{\bibfnamefont{A.}~\bibnamefont{Rasmussen}},
  \bibnamefont{and} \bibinfo{author}{\bibfnamefont{C.}~\bibnamefont{Xu}},
  \bibinfo{journal}{Phys. Rev. B} \textbf{\bibinfo{volume}{91}},
  \bibinfo{pages}{134404} (\bibinfo{year}{2015}).

\bibitem[{\citenamefont{Xu and Senthil}(2013)}]{xusenthil}
\bibinfo{author}{\bibfnamefont{C.}~\bibnamefont{Xu}} \bibnamefont{and}
  \bibinfo{author}{\bibfnamefont{T.}~\bibnamefont{Senthil}},
  \bibinfo{journal}{Phys. Rev. B} \textbf{\bibinfo{volume}{87}},
  \bibinfo{pages}{174412} (\bibinfo{year}{2013}).

\bibitem[{\citenamefont{Xu}(2013)}]{xu3dspt}
\bibinfo{author}{\bibfnamefont{C.}~\bibnamefont{Xu}}, \bibinfo{journal}{Phys.
  Rev. B} \textbf{\bibinfo{volume}{87}}, \bibinfo{pages}{144421}
  (\bibinfo{year}{2013}).

\bibitem[{\citenamefont{Witten}(1984)}]{Witten1984}
\bibinfo{author}{\bibfnamefont{E.}~\bibnamefont{Witten}},
  \bibinfo{journal}{Commun. Math. Phys.} \textbf{\bibinfo{volume}{92}},
  \bibinfo{pages}{455} (\bibinfo{year}{1984}).

\bibitem[{\citenamefont{Knizhnik and
  Zamolodchikov}(1984)}]{KnizhnikZamolodchikov1984}
\bibinfo{author}{\bibfnamefont{V.~G.} \bibnamefont{Knizhnik}} \bibnamefont{and}
  \bibinfo{author}{\bibfnamefont{A.~B.} \bibnamefont{Zamolodchikov}},
  \bibinfo{journal}{Nucl. Phys. B} \textbf{\bibinfo{volume}{247}},
  \bibinfo{pages}{83} (\bibinfo{year}{1984}).

\bibitem[{\citenamefont{Moon}(2012)}]{moonwzw}
\bibinfo{author}{\bibfnamefont{E.-G.} \bibnamefont{Moon}},
  \bibinfo{journal}{Phys. Rev. B} \textbf{\bibinfo{volume}{85}},
  \bibinfo{pages}{245123} (\bibinfo{year}{2012}).

\bibitem[{\citenamefont{Slagle et~al.}(2015)\citenamefont{Slagle, You, and
  Xu}}]{kevinQSH}
\bibinfo{author}{\bibfnamefont{K.}~\bibnamefont{Slagle}},
  \bibinfo{author}{\bibfnamefont{Y.-Z.} \bibnamefont{You}}, \bibnamefont{and}
  \bibinfo{author}{\bibfnamefont{C.}~\bibnamefont{Xu}}, \bibinfo{journal}{Phys.
  Rev. B} \textbf{\bibinfo{volume}{91}}, \bibinfo{pages}{115121}
  (\bibinfo{year}{2015}).

\bibitem[{\citenamefont{{He} et~al.}(2015)\citenamefont{{He}, {Wu}, {You},
  {Xu}, {Meng}, and {Lu}}}]{mengQSH2}
\bibinfo{author}{\bibfnamefont{Y.-Y.} \bibnamefont{{He}}},
  \bibinfo{author}{\bibfnamefont{H.-Q.} \bibnamefont{{Wu}}},
  \bibinfo{author}{\bibfnamefont{Y.-Z.} \bibnamefont{{You}}},
  \bibinfo{author}{\bibfnamefont{C.}~\bibnamefont{{Xu}}},
  \bibinfo{author}{\bibfnamefont{Z.~Y.} \bibnamefont{{Meng}}},
  \bibnamefont{and} \bibinfo{author}{\bibfnamefont{Z.-Y.} \bibnamefont{{Lu}}},
  \bibinfo{journal}{ArXiv e-prints}  (\bibinfo{year}{2015}),
  \eprint{1508.06389}.

\bibitem[{\citenamefont{Senthil
  et~al.}(2004{\natexlab{a}})\citenamefont{Senthil, Vishwanath, Balents,
  Sachdev, and Fisher}}]{deconfine1}
\bibinfo{author}{\bibfnamefont{T.}~\bibnamefont{Senthil}},
  \bibinfo{author}{\bibfnamefont{A.}~\bibnamefont{Vishwanath}},
  \bibinfo{author}{\bibfnamefont{L.}~\bibnamefont{Balents}},
  \bibinfo{author}{\bibfnamefont{S.}~\bibnamefont{Sachdev}}, \bibnamefont{and}
  \bibinfo{author}{\bibfnamefont{M.~P.~A.} \bibnamefont{Fisher}},
  \bibinfo{journal}{Science} \textbf{\bibinfo{volume}{303}},
  \bibinfo{pages}{1490} (\bibinfo{year}{2004}{\natexlab{a}}).

\bibitem[{\citenamefont{Senthil
  et~al.}(2004{\natexlab{b}})\citenamefont{Senthil, Balents, Sachdev,
  Vishwanath, and Fisher}}]{deconfine2}
\bibinfo{author}{\bibfnamefont{T.}~\bibnamefont{Senthil}},
  \bibinfo{author}{\bibfnamefont{L.}~\bibnamefont{Balents}},
  \bibinfo{author}{\bibfnamefont{S.}~\bibnamefont{Sachdev}},
  \bibinfo{author}{\bibfnamefont{A.}~\bibnamefont{Vishwanath}},
  \bibnamefont{and} \bibinfo{author}{\bibfnamefont{M.~P.~A.}
  \bibnamefont{Fisher}}, \bibinfo{journal}{Phys. Rev. B}
  \textbf{\bibinfo{volume}{70}}, \bibinfo{pages}{144407}
  (\bibinfo{year}{2004}{\natexlab{b}}).

\bibitem[{\citenamefont{Senthil and Fisher}(2005)}]{senthilfisher}
\bibinfo{author}{\bibfnamefont{T.}~\bibnamefont{Senthil}} \bibnamefont{and}
  \bibinfo{author}{\bibfnamefont{M.~P.~A.} \bibnamefont{Fisher}},
  \bibinfo{journal}{Phys. Rev. B} \textbf{\bibinfo{volume}{74}},
  \bibinfo{pages}{064405} (\bibinfo{year}{2005}).

\bibitem[{\citenamefont{Abanov}(2000)}]{abanov2}
\bibinfo{author}{\bibfnamefont{A.~G.} \bibnamefont{Abanov}},
  \bibinfo{journal}{Phys. Lett. B} \textbf{\bibinfo{volume}{321}},
  \bibinfo{pages}{492} (\bibinfo{year}{2000}).

\bibitem[{\citenamefont{Wang and Senthil}(2014)}]{senthilhe3}
\bibinfo{author}{\bibfnamefont{C.}~\bibnamefont{Wang}} \bibnamefont{and}
  \bibinfo{author}{\bibfnamefont{T.}~\bibnamefont{Senthil}},
  \bibinfo{journal}{Phys. Rev. B} \textbf{\bibinfo{volume}{89}},
  \bibinfo{pages}{195124} (\bibinfo{year}{2014}).

\bibitem[{\citenamefont{Hikami}(1980)}]{symmetricNLSM}
\bibinfo{author}{\bibfnamefont{S.}~\bibnamefont{Hikami}},
  \bibinfo{journal}{Prog. Theor. Phys.} \textbf{\bibinfo{volume}{64}},
  \bibinfo{pages}{1466} (\bibinfo{year}{1980}).

\bibitem[{\citenamefont{Brezin et~al.}(1980)\citenamefont{Brezin, Hikami, and
  Zinn-Justin}}]{generalizedNLSM}
\bibinfo{author}{\bibfnamefont{E.}~\bibnamefont{Brezin}},
  \bibinfo{author}{\bibfnamefont{S.}~\bibnamefont{Hikami}}, \bibnamefont{and}
  \bibinfo{author}{\bibfnamefont{J.}~\bibnamefont{Zinn-Justin}},
  \bibinfo{journal}{Nucl. Phys. B} \textbf{\bibinfo{volume}{165}},
  \bibinfo{pages}{528} (\bibinfo{year}{1980}).

\bibitem[{\citenamefont{Qi et~al.}(2008)\citenamefont{Qi, Hughes, and
  Zhang}}]{qi2008}
\bibinfo{author}{\bibfnamefont{X.-L.} \bibnamefont{Qi}},
  \bibinfo{author}{\bibfnamefont{T.~L.} \bibnamefont{Hughes}},
  \bibnamefont{and} \bibinfo{author}{\bibfnamefont{S.-C.} \bibnamefont{Zhang}},
  \bibinfo{journal}{Phys. Rev. B} \textbf{\bibinfo{volume}{78}},
  \bibinfo{pages}{195424} (\bibinfo{year}{2008}).

\bibitem[{\citenamefont{Halperin et~al.}(1974)\citenamefont{Halperin, Lubensky,
  and keng Ma}}]{halperincp}
\bibinfo{author}{\bibfnamefont{B.~I.} \bibnamefont{Halperin}},
  \bibinfo{author}{\bibfnamefont{T.~C.} \bibnamefont{Lubensky}},
  \bibnamefont{and} \bibinfo{author}{\bibfnamefont{S.}~\bibnamefont{keng Ma}},
  \bibinfo{journal}{Phys. Rev. Lett.} \textbf{\bibinfo{volume}{32}},
  \bibinfo{pages}{292} (\bibinfo{year}{1974}).

\bibitem[{\citenamefont{Kaul and Sachdev}(2008)}]{kaulsachdev}
\bibinfo{author}{\bibfnamefont{R.~K.} \bibnamefont{Kaul}} \bibnamefont{and}
  \bibinfo{author}{\bibfnamefont{S.}~\bibnamefont{Sachdev}},
  \bibinfo{journal}{Phys. Rev. B} \textbf{\bibinfo{volume}{77}},
  \bibinfo{pages}{155105} (\bibinfo{year}{2008}).

\bibitem[{\citenamefont{Fisher et~al.}(1972)\citenamefont{Fisher, keng Ma, and
  Nickel}}]{fisher1972}
\bibinfo{author}{\bibfnamefont{M.~E.} \bibnamefont{Fisher}},
  \bibinfo{author}{\bibfnamefont{S.}~\bibnamefont{keng Ma}}, \bibnamefont{and}
  \bibinfo{author}{\bibfnamefont{B.~G.} \bibnamefont{Nickel}},
  \bibinfo{journal}{Phys. Rev. Lett.} \textbf{\bibinfo{volume}{29}},
  \bibinfo{pages}{917} (\bibinfo{year}{1972}).

\bibitem[{\citenamefont{Dski and Kupianen}(1985)}]{GNrenormalizable}
\bibinfo{author}{\bibfnamefont{K.~G.} \bibnamefont{Dski}} \bibnamefont{and}
  \bibinfo{author}{\bibfnamefont{A.}~\bibnamefont{Kupianen}},
  \bibinfo{journal}{Nucl. Phys. B} \textbf{\bibinfo{volume}{262}},
  \bibinfo{pages}{33} (\bibinfo{year}{1985}).

\bibitem[{\citenamefont{Mross et~al.}(2010)\citenamefont{Mross, McGreevy, Liu,
  and Senthil}}]{senthilfermi}
\bibinfo{author}{\bibfnamefont{D.~F.} \bibnamefont{Mross}},
  \bibinfo{author}{\bibfnamefont{J.}~\bibnamefont{McGreevy}},
  \bibinfo{author}{\bibfnamefont{H.}~\bibnamefont{Liu}}, \bibnamefont{and}
  \bibinfo{author}{\bibfnamefont{T.}~\bibnamefont{Senthil}},
  \bibinfo{journal}{Phys. Rev. B} \textbf{\bibinfo{volume}{82}},
  \bibinfo{pages}{045121} (\bibinfo{year}{2010}).

\bibitem[{\citenamefont{Nayak and Wilczek}(1994{\natexlab{a}})}]{chetanfermi1}
\bibinfo{author}{\bibfnamefont{C.}~\bibnamefont{Nayak}} \bibnamefont{and}
  \bibinfo{author}{\bibfnamefont{F.}~\bibnamefont{Wilczek}},
  \bibinfo{journal}{Nucl. Phys. B} \textbf{\bibinfo{volume}{417}},
  \bibinfo{pages}{359} (\bibinfo{year}{1994}{\natexlab{a}}).

\bibitem[{\citenamefont{Nayak and Wilczek}(1994{\natexlab{b}})}]{chetanfermi2}
\bibinfo{author}{\bibfnamefont{C.}~\bibnamefont{Nayak}} \bibnamefont{and}
  \bibinfo{author}{\bibfnamefont{F.}~\bibnamefont{Wilczek}},
  \bibinfo{journal}{Nucl. Phys. B} \textbf{\bibinfo{volume}{430}},
  \bibinfo{pages}{534} (\bibinfo{year}{1994}{\natexlab{b}}).

\bibitem[{\citenamefont{Lee}(2009)}]{sslee}
\bibinfo{author}{\bibfnamefont{S.-S.} \bibnamefont{Lee}},
  \bibinfo{journal}{Phys. Rev. B} \textbf{\bibinfo{volume}{80}},
  \bibinfo{pages}{165102} (\bibinfo{year}{2009}).

\bibitem[{\citenamefont{Xu et~al.}(2008)\citenamefont{Xu, Mueller, and
  Sachdev}}]{xumuellersachdev}
\bibinfo{author}{\bibfnamefont{C.}~\bibnamefont{Xu}},
  \bibinfo{author}{\bibfnamefont{M.}~\bibnamefont{Mueller}}, \bibnamefont{and}
  \bibinfo{author}{\bibfnamefont{S.}~\bibnamefont{Sachdev}},
  \bibinfo{journal}{Phys. Rev. B} \textbf{\bibinfo{volume}{78}},
  \bibinfo{pages}{020501(R)} (\bibinfo{year}{2008}).

\bibitem[{\citenamefont{Frey and Balents}(1997)}]{freybalents}
\bibinfo{author}{\bibfnamefont{E.}~\bibnamefont{Frey}} \bibnamefont{and}
  \bibinfo{author}{\bibfnamefont{L.}~\bibnamefont{Balents}},
  \bibinfo{journal}{Phys. Rev. B} \textbf{\bibinfo{volume}{55}},
  \bibinfo{pages}{1050} (\bibinfo{year}{1997}).

\bibitem[{\citenamefont{Wen}(2003)}]{wenphoton}
\bibinfo{author}{\bibfnamefont{X.-G.} \bibnamefont{Wen}},
  \bibinfo{journal}{Phys. Rev. B} \textbf{\bibinfo{volume}{68}},
  \bibinfo{pages}{115413} (\bibinfo{year}{2003}).

\bibitem[{\citenamefont{Moessner and Sondhi}(2003)}]{sondhiphoton}
\bibinfo{author}{\bibfnamefont{R.}~\bibnamefont{Moessner}} \bibnamefont{and}
  \bibinfo{author}{\bibfnamefont{S.~L.} \bibnamefont{Sondhi}},
  \bibinfo{journal}{Phys. Rev. B} \textbf{\bibinfo{volume}{68}},
  \bibinfo{pages}{184512} (\bibinfo{year}{2003}).

\bibitem[{\citenamefont{Hermele et~al.}(2004)\citenamefont{Hermele, Fisher, and
  Balents}}]{hermelephoton}
\bibinfo{author}{\bibfnamefont{M.}~\bibnamefont{Hermele}},
  \bibinfo{author}{\bibfnamefont{M.~P.~A.} \bibnamefont{Fisher}},
  \bibnamefont{and} \bibinfo{author}{\bibfnamefont{L.}~\bibnamefont{Balents}},
  \bibinfo{journal}{Phys. Rev. B} \textbf{\bibinfo{volume}{69}},
  \bibinfo{pages}{064404} (\bibinfo{year}{2004}).

\bibitem[{\citenamefont{Arkani-Hamed et~al.}(2001)\citenamefont{Arkani-Hamed,
  Cohen, and Georgi}}]{littlehiggs}
\bibinfo{author}{\bibfnamefont{N.}~\bibnamefont{Arkani-Hamed}},
  \bibinfo{author}{\bibfnamefont{A.~G.} \bibnamefont{Cohen}}, \bibnamefont{and}
  \bibinfo{author}{\bibfnamefont{H.}~\bibnamefont{Georgi}},
  \bibinfo{journal}{Phys. Rev. B} \textbf{\bibinfo{volume}{513}},
  \bibinfo{pages}{232} (\bibinfo{year}{2001}).

\bibitem[{\citenamefont{Arkani-Hamed
  et~al.}(2002{\natexlab{a}})\citenamefont{Arkani-Hamed, Cohen, Katz, Nelson,
  Gregoire, and Wacker}}]{littlehiggs1}
\bibinfo{author}{\bibfnamefont{N.}~\bibnamefont{Arkani-Hamed}},
  \bibinfo{author}{\bibfnamefont{A.~G.} \bibnamefont{Cohen}},
  \bibinfo{author}{\bibfnamefont{E.}~\bibnamefont{Katz}},
  \bibinfo{author}{\bibfnamefont{A.~E.} \bibnamefont{Nelson}},
  \bibinfo{author}{\bibfnamefont{T.}~\bibnamefont{Gregoire}}, \bibnamefont{and}
  \bibinfo{author}{\bibfnamefont{J.~G.} \bibnamefont{Wacker}},
  \bibinfo{journal}{JHEP} \textbf{\bibinfo{volume}{0208}}, \bibinfo{pages}{019}
  (\bibinfo{year}{2002}{\natexlab{a}}).

\bibitem[{\citenamefont{Arkani-Hamed
  et~al.}(2002{\natexlab{b}})\citenamefont{Arkani-Hamed, Cohen, Katz, and
  Nelson}}]{littlehiggs2}
\bibinfo{author}{\bibfnamefont{N.}~\bibnamefont{Arkani-Hamed}},
  \bibinfo{author}{\bibfnamefont{A.~G.} \bibnamefont{Cohen}},
  \bibinfo{author}{\bibfnamefont{E.}~\bibnamefont{Katz}}, \bibnamefont{and}
  \bibinfo{author}{\bibfnamefont{A.~E.} \bibnamefont{Nelson}},
  \bibinfo{journal}{JHEP} \textbf{\bibinfo{volume}{0207}}, \bibinfo{pages}{034}
  (\bibinfo{year}{2002}{\natexlab{b}}).

\bibitem[{\citenamefont{Kaplan and Schmaltz}(2003)}]{littlehiggs3}
\bibinfo{author}{\bibfnamefont{D.~E.} \bibnamefont{Kaplan}} \bibnamefont{and}
  \bibinfo{author}{\bibfnamefont{M.}~\bibnamefont{Schmaltz}},
  \bibinfo{journal}{JHEP} \textbf{\bibinfo{volume}{0310}}, \bibinfo{pages}{039}
  (\bibinfo{year}{2003}).

\end{thebibliography}

\end{document}